\newcommand{\al}{\alpha}

\newcommand{\Ma}{M_{\ast}}

\input{aipcheck}

\documentclass[
   ,draft            
  ,numberedheadings 
  ]
  {aipproc}

\usepackage{amssymb}
\usepackage{amsmath}
\usepackage[all]{xy}

\newcommand{\ba}{\begin{eqnarray}}
\newcommand{\ea}{\end{eqnarray}}
\newcommand{\be}{\begin{equation}}
\newcommand{\ee}{\end{equation}}

\layoutstyle{6x9}

\begin{document}

\title{Cosmology of \\ Modified (but second order) Gravity}


\author{Tomi S. Koivisto}{
  address={Institute f\"ur theoretische Physik, Heidelberg Universit\"at, \\ Philosophenweg 16, 061920 Heidelberg, Deutschland; \\ 
           Institute for Theoretical Physics and Spinoza Institute, Utrecht University, \\
           Leuvenlaan 4, Postbus 80.195, 3508 TD Utrecht, The Netherlands}
}

\begin{abstract}

This is a brief review of applications of extended gravity theories to cosmology, in particular to the dark energy problem.
Generically extensions of gravity action involve higher derivative terms, which can result in ghosts and instabilities.
We consider three ways to circumvent this: Chern-Simons terms, first order variational principle and nonlocality. We consider 
some recent cosmological applications of these three classes of modified gravity models.
The viable parameter spaces can be very efficiently bounded by taking into account cosmological constraints from all epochs in addition to 
bounds from Solar system tests and stability considerations. We make some new remarks concerning so called algebraic scalar-tensor theories,
biscalar reformulation of nonlocal actions involving the inverse d'Alembertian, and a possible covariant formulation holographic cosmology with
nonperturbative gravity.
 
\end{abstract}

\maketitle


\section{Introduction}


General theory of relativity (GR) has to be extended at high energy scales to resolve its singularities, the hierarchy problem, quantization and 
unification with other theories of fundamental interactions. The dark energy energy problem \cite{Copeland:2006wr} has raised speculations about the possibility of an infrared modification of GR too. Alternative theories of gravitation 
have been around since Nordstr\"oms scalar gravitation \cite{Nordstrom:1988fi}, but no evidence for departure from the classical predictions of GR have been observed. One may then contemplate if the
acceleration of the universe (and perhaps also the need to invoke dark matter) could be a sign of a need to modify GR. 

In Uzan's review of possible physics behind the cosmic acceleration \cite{Uzan:2006mf}, a classification of different approaches 
was based on which kind of new fields they introduce and how these new fields couple to the matter and the metric. Generically, modifying
gravity is equivalent to invoking a new matter field. In the simplest case of a Brans-Dicke -like theory \cite{Brans:1961sx} it is a scalar field but
other types of fields can appear as well, for example tensors in bimetric theories. Usually also several fields come about, like in
low-energy limit of string theory featuring typically form fields, the dilaton scalar field and other moduli. 

In the present discussion we will focus on models which {\it avoid} introducing new fields while modifying gravitational interactions.
The motivation for this is that, as will be explained in the next section, rather generically the new fields would actually behave pathologically and thus actually render the starting point invalid.
It is interesting to contemplate how well such specific classes of gravity theories are constrained, since in principle all other theories can be understood as new matter in disguise (as far as only gravity 
is concerned). There is also some appeal in the attempts to deal away dark energy without invoking any fields, though of course no fundamental principle excludes the possibility of the dark sector being arbitrarily 
complicated: indeed it could involve as many fields as the visible sector described by the Standard model - or even vast number of more fields \cite{Dvali:2009fw}. 
 
This starting point complements the reviews of modified gravity cosmology \cite{Nojiri:2006ri,Capozziello:2007ec,Sotiriou:2008rp,Lobo:2008sg,Sami:2009dk}.  
The plan of the presentation is as follows. In Section \ref{ext} we present simple field theory considerations which provide the basis for our discussion. The lesson is that unless new fields were introduced, 
only specific classes of effective field theories are viable. We then look at this within the scalar-tensor framework. Then three classes of models are described and a brief summary of their cosmological
applications to the dark energy problem is provided: the Gauss-Bonnet model in Section \ref{gb-sec}, the first order formulation of modified gravity in Section \ref{pa-sec} and nonlocal gravity models
in Section \ref{nl-sec}. We conclude in Section \ref{co-sec}. 
  
\section{Extensions the gravity action}
\label{ext}

Consider a generic Lagrangian involving derivatives to order $N$, $L(q,\dot{q},\dots,q^{(N)})$. 
The canonical phase space then has $2N$ coordinates\footnote{We assume the Lagrangian depends nondegenerately upon $q^{(N)}$.}, and they may be defined as
\be
Q_i \equiv q^{(i-1)}\,, \quad P_i \equiv \sum^N_{j=1}\left(-\frac{d}{dt}\right)^{j-i}\frac{\partial L}{\partial q^{(j)}}\,.
\ee
One might use the equation of motion to solve $q^{(N)}=A(Q_1,\dots,Q_N,P_N)$ in terms of the other variables. However, the other degrees of freedom remain unconstrained and contribute to the Hamiltonian with indefinite sign:
\be
H \equiv \sum_{i=1}^{N}P_iq^{(i)}-L=P_1 Q_2 + P_2 Q_3 +\dots + P_{N-1}Q_N + P_N A-L(Q_1,\dots,Q_N,A)\,.
\ee
Thus the Hamiltonian is unbounded from below. This result is generic, unless additional constrains are introduced \cite{Simon:1990ic}, higher derivative theories lack a minimum energy state.
This may be called the Ostrogradski theorem. See Woodard's lecture \cite{Woodard:2006nt} for a lucid presentation of the argument and discussion within the framework of generalized gravity theories. 

Below we will consider a consider prototype of generalized gravity, scalar-tensor theory, from the point of view of eliminating or constraining the additional degree of freedom.  
There is a one-parameter family of classes of scalar-tensor theories in which the scalar field is given by an
algebraic function of the curvature and the trace of matter content. The
family of these theories provides a new perspective to the nonlinear gravity theories, since it includes both their
metric and Palatini formulations as special cases. It is then possible to continuously interpolate between these
formulations and avoid some pathologies. To end this section, we resume in \ref{tre} the possible approaches to modify gravity avoiding the Ostrogradskian catastrophy. 

\subsection{A scalar-tensor viewpoint}

We consider the scalar-tensor action in $n$ dimensions
\be \label{action}
S = \int d^n x \sqrt{-g}\left[\frac{1}{2}F(\phi)R - \frac{1}{2}\omega(\phi)(\partial\phi)^2 - V(\phi) + W(\phi)\mathcal{L}_{(m)}\right],
\ee
where $\mathcal{L}_{(m)}$ is the matter Lagrangian and the corresponding stress energy tensor is defined as 
 \be \label{memt}
 T_{\mu\nu} \equiv -\frac{2}{\sqrt{-g}} \frac{\delta
 (\sqrt{-g}\mathcal{L}_m)}{\delta(g^{\mu\nu})}.
 \ee
The field equation for the metric follows as
\ba \label{mfield}
F(\phi) G_{\mu\nu} & = & \left(\nabla_\mu\nabla_\nu - g_{\mu\nu}\Box \right) F(\phi)
+ \omega(\phi)\left[(\nabla_\mu\phi)(\nabla_\nu\phi)-\frac{1}{2}(\partial\phi)^2g_{\mu\nu}\right] \nonumber \\ & - & V(\phi)g_{\mu\nu} + W(\phi)T^{(m)}_{\mu\nu}.
\ea
The scalar field obeys the equation of motion
\be \label{sfield}
\omega(\phi)\Box \phi + \frac{1}{2}\omega'(\phi)(\partial\phi)^2 - V'(\phi) + \frac{1}{2}F'(\phi)R = -W'(\phi)\mathcal{L}_{(m)}.
\ee
From these equations one may infer that the matter sector obeys
\be \label{matter}
\nabla^\mu T^{(m)}_{\mu\nu} = \left(\nabla^\mu\log{W(\phi)}\right)\left(g_{\mu\nu}\mathcal{L}^{(m)}-T^{(m)}_{\mu\nu}\right)\,,
\ee
expressing the covariant energy-momentum conservation \cite{Koivisto:2005yk}.

\subsubsection{The algebraic family}

By combining the trace of the field equation (\ref{mfield}) and the generalized Klein-Gordon equation for the scalar field (\ref{sfield}),
the scalar curvature can be expressed as
\ba
\left[(1-\frac{n}{2})F-\frac{1}{2}(n-1)\frac{F'^2}{\omega}\right]R & = &
\left[(1-\frac{n}{2})\omega +\frac{1}{2}(n-1)\omega'\frac{F'}{\omega} - (n-1)F''\right](\partial \phi)^2 \nonumber \\
& - & nV-(n-1)\frac{F'}{\omega}V' + WT_{(m)} +
(n-1)\frac{F'}{\omega}W'\mathcal{L}_{(m)}.
\ea
We ease notation, we will occasionally drop the explicit $\phi$-dependence from the functions as we did here.
If the first square bracket term in the LHS vanishes, there is an algebraic relation between the curvature, the matter variables and the scalar field.
Otherwise such a unique relation does not exist, though it is satisfied by any theory in the dynamical situations where the derivatives of the field
happen to vanish. If the relation holds universally within the theory, we call it {\it an algebraic scalar-tensor theory}.

To analyse their properties, it is convenient to set $F(\phi)=\phi$. We can do this without any essential loss of generality, since as long as $F(\phi)$
is invertible, the change is equivalent to a redefinition of $\omega$, $V$ and $W$. Now it becomes easy to show that the algebraic theories allow only a certain
form of the kinetic term. In particular, the requirement is that
\be \label{kinetic}
\omega(\phi) = \frac{\Omega_n}{\Omega_A-\phi},
\ee
where $\Omega_A$ is an arbitrary constant (stemming from an integration constant), and the constant $\Omega_n$ depends on the dimensionality as
\be \label{omega}
\Omega_n = \frac{n-1}{n-2}.
\ee
The constant $\Omega_A \in \mathbf{R}$ is allowed to take arbitrary values.
Thus the algebraic theories constitute a one-parameter family of classes of theories
in the space of all possible scalar-tensor theories.

The nonlinear gravity models, the so called $f(R)$ gravities, are specific examples of these algebraic theories.
In the metric formulation of nonlinear gravity, the kinetic term is absent and thus this corresponds to the limit
of infinite $\Omega_A$. The field is still dynamical, because it is nonminimally coupled to gravity.
However, it can be expressed a function of a $R$ alone, $\phi=\phi(R)$. It has been argued that indeed in this particular case one may succeed in avoiding the Ostrograskian instabilities while
increasing the order of derivatives \cite{Woodard:2006nt}. However, in the present discussion we do not go into details of this case. Extensive studies are found in the literature
(for a list of reviews, see the introduction).

The case corresponding to the limit of vanishing $\Omega_A$ is nothing but the ''$\omega=-3/2$ Brans-Dicke theory'' and it is in many regards the 
opposite of the metric $f(R)$ models. 
This is  equivalent\footnote{Or slightly more general, since the potential can be arbitrary and does not have to be obtained from a particular function $f(R)$.} to the first order formulation of nonlinear gravity (i.e. Palatini-$f(R)$ models) which we discuss in more detail in section \ref{pa-sec}. In such a scalar-tensor theory 
then the field is not dynamical, since the kinetic term exactly cancels the effect of nonminimal gravity coupling. One may express the field as a function of $T$ alone, $\phi=\phi(T)$.

As one interpolates between the extremes $\Omega_A=0$ and $\Omega_A\rightarrow\infty$, the scalar field always depends on the both the trace of matter and the
Ricci curvature. However, these theories become equivalent to metric nonlinear gravity in vacuum or in the presence of conformal
matter; and they are equivalent to Palatini nonlinear gravity in Ricci-flat spacetimes.
The algebraic scalar-tensor theories are conformally coupled scalar models, whose the gravitational constant is given by $\Omega_A$. 

\subsection{Three approaches}
\label{tre}

There are three frameworks of extending gravity theories without introducing new degrees of freedom,
in particular avoiding ghosts.
\begin{itemize}
\item Gauss-Bonnet models. The Gauss-Bonnet invariant $G = R^2-4R_{\mu\nu}R^{\mu\nu}+R_{\mu\nu\rho\sigma}R^{\mu\nu\rho\sigma}$
combines the quadratic invariants in such a way, unique in four dimensions, that the higher derivative contributions cancel.
Models based on this term could thus avoid higher derivatives.
\item Palatini models. In the Palatini framework one regards connection and the metric as independent variables. This means working on
first order formalism instead of the second order one, or in Hamiltonian instead of Lagrangian formulation. Then one can consider extended
action without necessarily increasing the order of the theory.
\item Nonlocal models. For an action involving infinite number of derivatives, the Ostrogradski-type argument above can break down
as a truncation to cannot be performed at any finite order to construct the canonical momentum. Such nonlocal models can modify the
gravitational interaction without new propagating modes.
\end{itemize}
Gravitational alternatives to dark energy belonging to each of these frameworks have received considerable attentation in the literature
during few recent years. In the following we review concisely what has been learned about their cosmological implications, not aiming at an 
exhaustive account, but a summary of the main features of some presentative examples.


\section{Gauss-Bonnet models}
\label{gb-sec}

The action for the system to consider is
\be
S = \int d^4 \sqrt{-g}\left[\frac{M^2}{2}R-
\frac{\gamma}{2}(\nabla\phi)^2 - V(\phi) -f(\phi)G+L_m\right]
\ee
where $\gamma$ is a constant and $f(\phi)$ is the coupling which appears in the one-loop corrected string effective Einstein frame action \cite{Antoniadis2,Antoniadis3}.
The Gauss-Bonnet term $G$ is topological and thus doesn't alone contribute to the field equations. Therefore we cheat by including the field $\phi$, which 
can be identified with the string theory dilaton or a modulus.
We parameterize the coupling and the potential as
\be V(\phi) = V_0e^{\lambda\phi/(\sqrt{2}M)}, \quad
f(\phi) = f_0e^{\alpha\phi/(\sqrt{2}M)}\,. \label{param}
\ee
The nonperturbative effects from gaugino condensation or instanton can result an exponential potential. An exponential field dependence of the coupling is known to occur in supergravity actions. 
These forms are thus theoretically well motivated. We will see that the exponential case happens to have some beneficial features for phenomenology too.

\subsection{Gauss-Bonnet models: Cosmology}

Though the severe constraints on Gauss-Bonnet coupling of dark energy were mentioned in \cite{EspositoFarese:2003ze,EspositoFarese:2004cc}, the possibility of Gauss-Bonnet 
interaction of dark energy has been considered in several studies \cite{Nojiri:2005vv,Calcagni:2005im,Carter:2005fu,Neupane:2006dp}.
Here, in particular we discuss the scenario proposed in \cite{Koivisto:2006xf}, see also \cite{Tsujikawa:2006ph}. Other models include $f(G)$ gravity \cite{Nojiri:2005jg}, generalized couplings 
\cite{Amendola:2005cr}, vector couplings \cite{Koivisto:2008xf} and brane-world models \cite{Zhang:2009dw}.  
 
We consider the flat Friedmann-Lemaitre-Robertson-Walker (FLRW) metric
\be \label{metric}
ds^2 = -dt^2 + a^2(t)dx^2
\ee 
and refer to $H \equiv \dot{a}/a$ as the Hubble parameter, an overdot meaning derivative with respect to $t$.
The equations of motion for the scale factor and the scalar field are then
\be \label{gb_eom1}
3M^2H^2 = \frac{\gamma}{2}\dot{\phi}^2 + a^2V(\phi) +
24a^{-2}H^3f'(\phi)\phi + \rho_m\,,
\ee
\be \label{gb_eom2}
\gamma(\ddot{\phi}+2H\dot{\phi})+a^2V'(\phi)+24a^{-2}f'(\phi)H^2\dot{H}=0\,.
\ee
For the purposes of phase space analysis, it is convenient to introduce the dimensionless variables defined as
\ba \label{variables}
\Omega_m &\equiv& \frac{\kappa^2\rho_m}{3H^2}, \quad
x \equiv \frac{\kappa}{\sqrt{2}}\frac{\dot{\phi}}{H}, \quad
y \equiv \kappa^2\frac{V(\phi)}{H^2}, \nonumber \\
\mu  &\equiv& 8\kappa^2\dot{\phi} H f'(\phi), \quad
w_{eff} \equiv -\frac{3}{2}\frac{\dot{H}}{H^2}-1.
\ea
The equations (\ref{gb_eom1},\ref{gb_eom2}) may then be rewritten in terms of such variables \cite{Neupane:2006dp}, and the fixed points of the dynamical system that
we find corresponding to the exponential model (\ref{param}) are then described in the Table \ref{tb2}. The stability properties of these fixed points are summarized in Table \ref{tb1}, for
the details see Ref.\cite{Koivisto:2006ai}. The dark energy scenario is basically the transition from the scaling fixed point E to the accelerated fixed point G, as we describe below.

\begin{table}
\begin{tabular}{|c|c|c|c|c|c|}
\hline
Fixed point & $x$ & $y$ & $\mu$ & $w_{eff}$ \\ \hline
A & 0 & 0 & $0$ & $w_m$ \\ \hline
B & $\pm 3$ & $0$ & $0$ & $1$ \\ \hline
C & $0$ & $3$ & 0 & $-1$ \\ \hline
D & $\lambda/2$ & $3-\lambda^2/4$ &0 & $\lambda^2/6-1$ \\ \hline
E & $3(1+w_m)/\lambda$ & $9(1-w_m^2)/\lambda^2$ & $0$ & $w_m$ \\ \hline
F & $3(1+w_m)/\alpha$ & $0$ & $18\frac{1-w_m}{1+3w_m}(1+w_m)^2$ & $w_m$ \\ \hline
G & $\neq 0$ & $0$ & $\neq 0$ & $w$ \\ \hline
\end{tabular}
\caption{\label{tb2} The critical points in the system.}
\end{table}

\begin{table}
\begin{tabular}{|c|c|c|}
\hline
Fixed point & stable?  & description \\ \hline
A & no &  matter domination \\ \hline
B & no &  kination  \\ \hline
C & if $\alpha > \lambda$ & de Sitter \\ \hline
D & doesn't exist for $\lambda > 2\sqrt{3}$ & scalar dominated \\ \hline
E & saddle point if $\alpha > \lambda$ & scaling \\ \hline
F & no & G-B scaling \\ \hline
G & no & G-B domination \\ \hline
\end{tabular}
\caption{\label{tb1} Description of the critical points in the system and a comment on their stability or existence.}
\end{table}

\begin{figure}
  \includegraphics[height=.26\textheight]{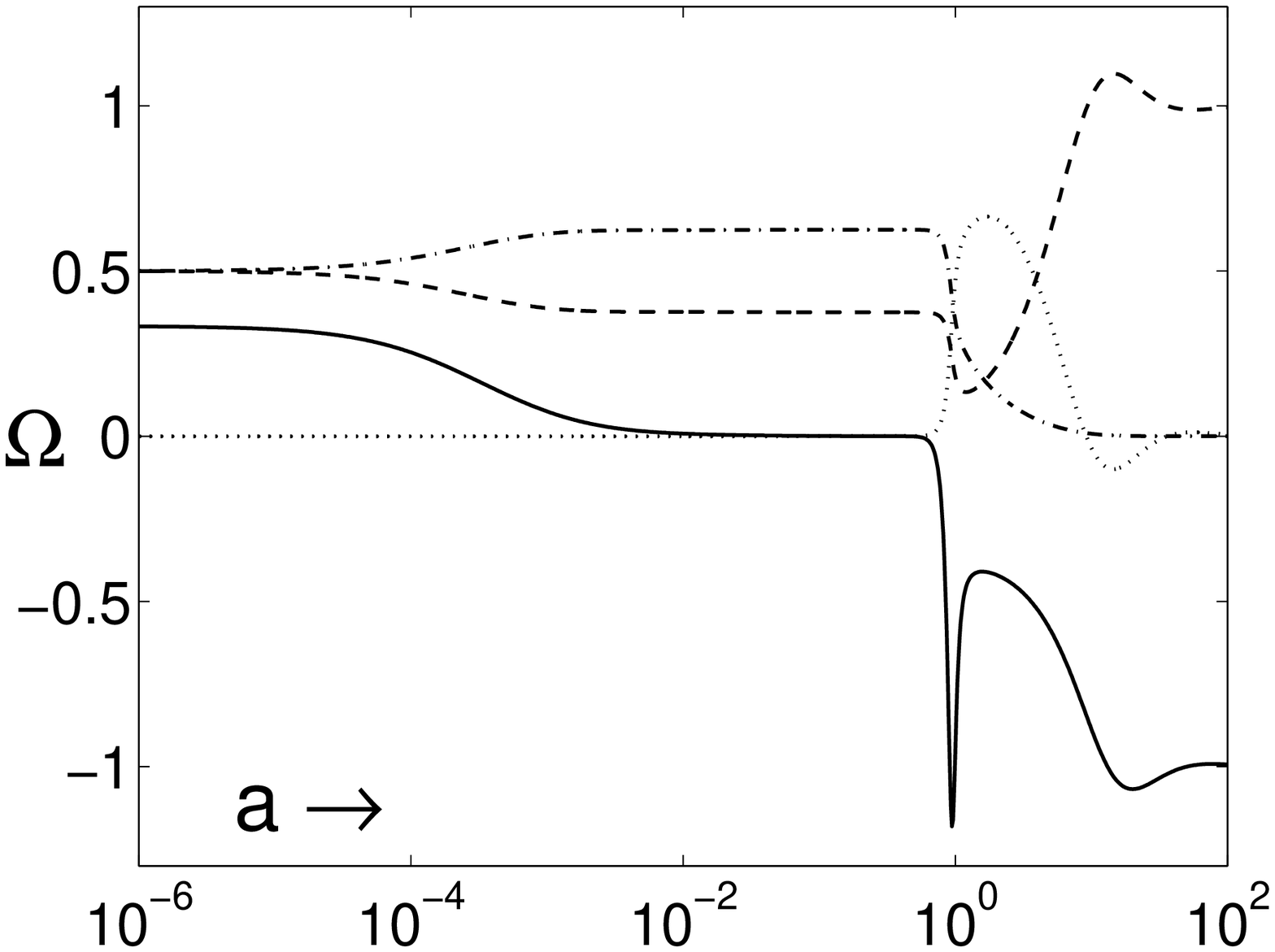}
  \includegraphics[height=.26\textheight]{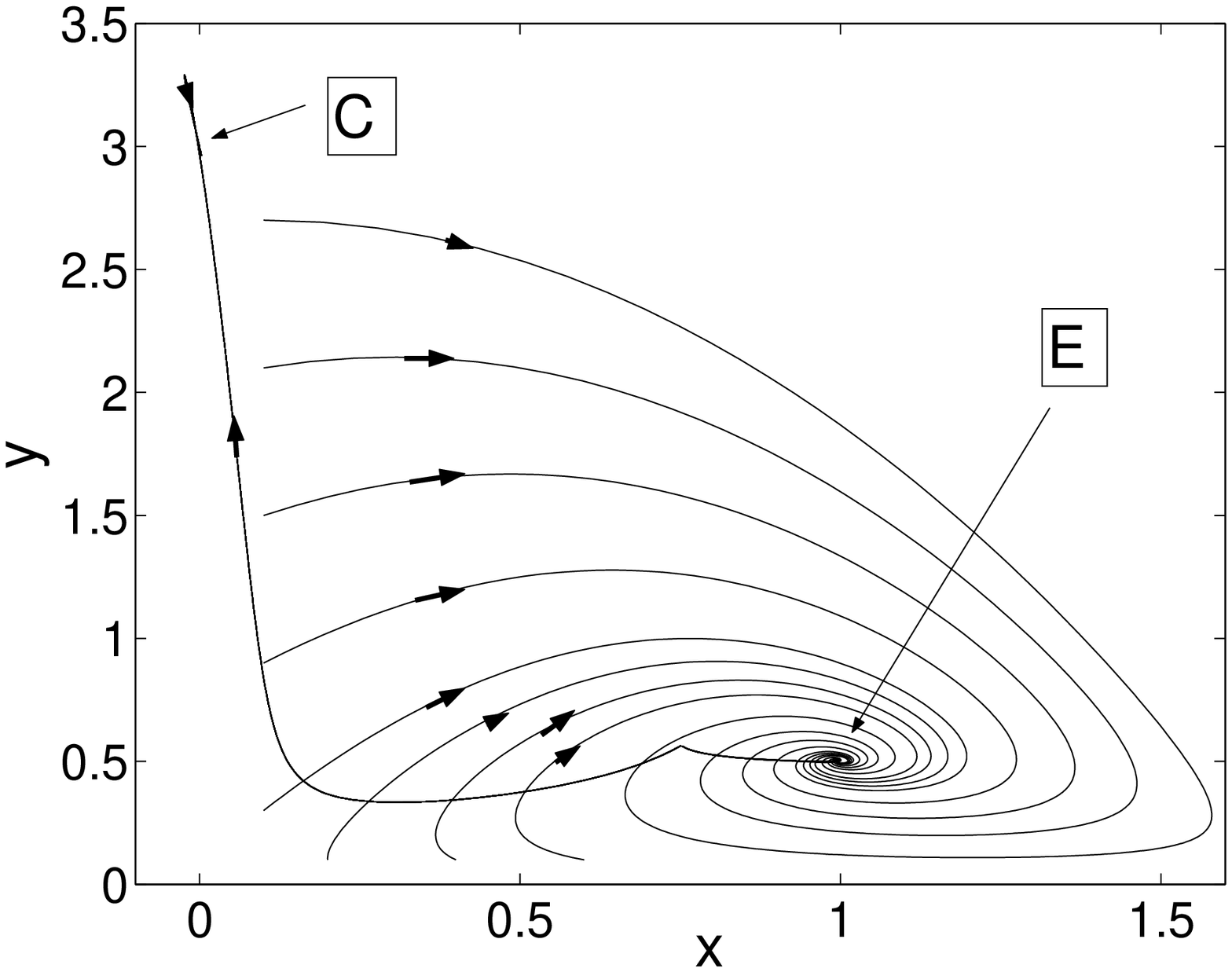}
  \caption{Left panel depicts the fractional energy densities
for matter, $\Omega_m$ (dash-dotted line), the scalar field
$\Omega_\phi$ (dashed line) and Gauss-Bonnet
term, $\Omega_f$ (dotted line). The solid line is the total
equation of state $w_{eff}$. 
and the transient phantom era is caused by the dynamics of the coupling.
Right panel is a portrait of the phase space. Here $\lambda=4$ and $\alpha=20$. 
\label{gb_fig}}
\end{figure}

The exponential quintessence is known to mimic the background dominating component equation of state. This so called scaling solution
is an attractor, and thus the evolution of the field is widely independent of the initial conditions \cite{Copeland:1997et}. To accelerate the universe however, the field should be tossed of the attractor. Now this is naturally done by the 
Gauss-Bonnet coupling, if $\alpha>\lambda$.
Given this simple condition, the scaling attractor becomes a saddle point from which the universe enters into
the asymptotically de Sitter space. Possible transient effects include
phantom-like expansion with $w<-1$. The full evolution of the universe
is shown in the figure \ref{gb_fig}.

It is pleasing that not only the form of the model but also the numerical values relevant for this scenario can be motivated by low energy string actions, since the parameters we need for 
the present acceleration in (\ref{param}) have natural scales. The dimensionless parameters $\al$ and $\lambda$ are of the
order one, and due to the shift symmetry of the potential,
one can always consider $V_0 \sim M^4$ and $f_0 \sim M^{-4}$. The initial conditions of the field are (almost) arbitrary. Similar exploitation of the exponential form can be deviced by adjusting the
form of the potential of the field or disforming its coupling to the metric (when in fact a similar transition mechanism emerges \cite{Koivisto:2008ak}).   

The Solar system limit of these models is problematical, as already observed by Esposito-Farese \cite{EspositoFarese:2003ze,EspositoFarese:2004cc}. The Gauss-Bonnet term behaves as
$r^{-4}$, and it is then difficult to neglect its small scale effects. Recently ways out have been considered, by reconstructing couplings that reproduce GR at small scales \cite{Amendola:2008vd} or
applying the chameleon mechanism \cite{Ito:2009nk}. At the cosmological scales new effects may occur. The evolution equation for the matter overdensity $\delta$ is given by 
\be \label{d_evol}
\ddot{\delta} + 2H\dot{\delta} = 4\pi G_*\rho \delta.
\ee
The effective gravitational constant $G_*$ seen by the matter inhomogeneities
depends rather non-trivially on the evolution of the background quantities, and in terms of our
dimensionless variables defined in Eq.(\ref{variables}). It can then be written as  
\be \label{g_eff}
G_* = 4\frac{-x^4 + \mu^2(1+\epsilon)^2 +
        x^2[2(1 + \epsilon)(\mu-1) + y]}{x^2[4 + \mu(5\mu -8)] - \mu^2[6(1 + \epsilon)(\mu-1) + y]}
\ee
In fact, it is possible that the expression for the denominator vanishes, and the effective gravitational constant diverges. This is linked with an appearance of a 
ghost \cite{Koivisto:2006ai}. Such peculiar instabilities exist both for tensor and scalar perturbations in one-loop string cosmology. Some of these instabilities were first 
discovered in the pre-big bang cosmology \cite{Kawai:1998ab}. Recent considerations of stability include \cite{Calcagni:2006ye,DeFelice:2006pg,Cognola:2007vq,DeFelice:2009wp}. 

\section{Palatini models}
\label{pa-sec}
                                                                                                    
Consider gravity theories represented by the action\footnote{More general functions $f(g^{\mu\nu},\hat{R}_{\mu\nu})$ can be considered too \cite{Allemandi:2004wn,Olmo:2009xy,Li:2008fa}, but we restrict 
here to the simplest case. In addition, if matter was nonminimally coupled to the independent connection we would be lead to more general models characterized by nonzero hypermomentum \cite{Sotiriou:2006qn}.
For our purposes of demonstrating cosmological phenomenology the simplest case is enough, however to address the criticisms concerning the validity of the averaging 
procedure \cite{Li:2008bma}, existence of simplest star solutions \cite{Barausse:2007pn} and well-posedness of the Cauchy problem \cite{LanahanTremblay:2007sg},
one may be forced into a more general framework.}  
 \be \label{action_p}
 S = \int d^4x \sqrt{-g}
    \left[\frac{1}{2}f(g^{\mu\nu}\hat{R}_{\mu\nu})
    + \mathcal{L}_m(g_{\mu\nu},\phi,...)\right].
 \ee
Here $\phi,...$ are some matter fields. In the Palatini approach one lets the torsionless connection
$\hat{\Gamma}^\alpha_{\beta\gamma}$ vary independently of the metric. The Ricci tensor is then constructed
solely from this connection,
 \be \label{ricci}
 R_{\mu\nu} \equiv \hat{\Gamma}^\alpha_{\mu\nu , \alpha}
       - \hat{\Gamma}^\alpha_{\mu\alpha , \nu}
       + \hat{\Gamma}^\alpha_{\alpha\lambda}\hat{\Gamma}^\lambda_{\mu\nu}
        - \hat{\Gamma}^\alpha_{\mu\lambda}\hat{\Gamma}^\lambda_{\alpha\nu}\,.
 \ee
In the following we will call $R \equiv g^{\mu\nu}\hat{R}_{\mu\nu}$. 
The field equations which follow from extremization of the action,
Eq.(\ref{action_p}), with respect to metric variations, can be
written as
 \be \label{fields2} F R^\mu_\nu -\frac{1}{2}f\delta^\mu_\nu =
 T^\mu_\nu \,,
 \ee
where we have defined $F \equiv \partial f/\partial R$.  In GR, $f=\kappa^2(R-2\Lambda)$, so $F = \kappa^2$. By varying the action with respect to $\hat{\Gamma}^\alpha_{\beta\gamma}$, one gets the
condition
 \be
 \hat{\nabla}_\alpha\left[\sqrt{-g}g^{\beta\gamma}F\right]=0,
 \ee
implying that this connection is compatible with the conformal metric
 \be
 \hat{g}_{\mu\nu} \equiv F^{2/(n-2)}g_{\mu\nu} \,.
 \ee
This connection governs how the tensor $R_{\mu\nu}$ appearing in the
action settles itself in order to minimize the action, but it turns
out that the metric compatible connection determines the geodesics
that freely falling particles follow, since the energy momentum defined as (\ref{memt})
is conserved according to this connection \cite{Koivisto:2005yk},
 $\nabla_\mu T^\mu_\nu = 0$,
whereas in general $\hat{\nabla}_\mu T^\mu_\nu \neq 0$. Therefore one
may interpret the Levi-Civita connection as the gravitational field
as in Ref.\cite{Magnano:1995pv}. The trace of the field equations allows
us to solve $R$ as an algebraic function of the matter trace $T \equiv g^{\mu\nu}T_{\mu\nu}$. This central relation is
 \be \label{trace} FR-2f = T\,.
 \ee
Written in the form of GR plus correction terms, the field equations read:
 \ba \label{eg} G^\mu_\nu(g) & = & {T^\mu_\nu}
               + (1-F)R^\mu_\nu(g)
               -  \frac{3}{2F}(\nabla^\mu F)(\nabla_\nu F) \nonumber \\ \label{p_fe}
               +   \nabla^\mu \nabla_\nu F
             &  + & \frac{1}{2}\left[(f-R) + (1-\frac{3}{F})\Box F +
     \frac{3}{2F}(\partial F)^2\right]\delta^\mu_\nu.
 \ea
Since the corrections can be expressed as functions of the matter trace, one can view Eq.(\ref{eg}) as GR with nonstandard matter couplings\footnote{An unusual approach with possibly similar consequences 
is to consider $f(R,T)$ Lagrangians \cite{Poplawski:2006ey}.}. 
The whole RHS may be regarded as an effective matter energy-momentum tensor. In vacuum it reduces to a cosmological constant \cite{Ferraris:1992dx}. This is also the case in the presence of conformal
matter, i.e. if $T=0$.

\subsection{Palatini models: Cosmology}
\label{pala_struc}

\begin{figure}
  \includegraphics[height=.26\textheight]{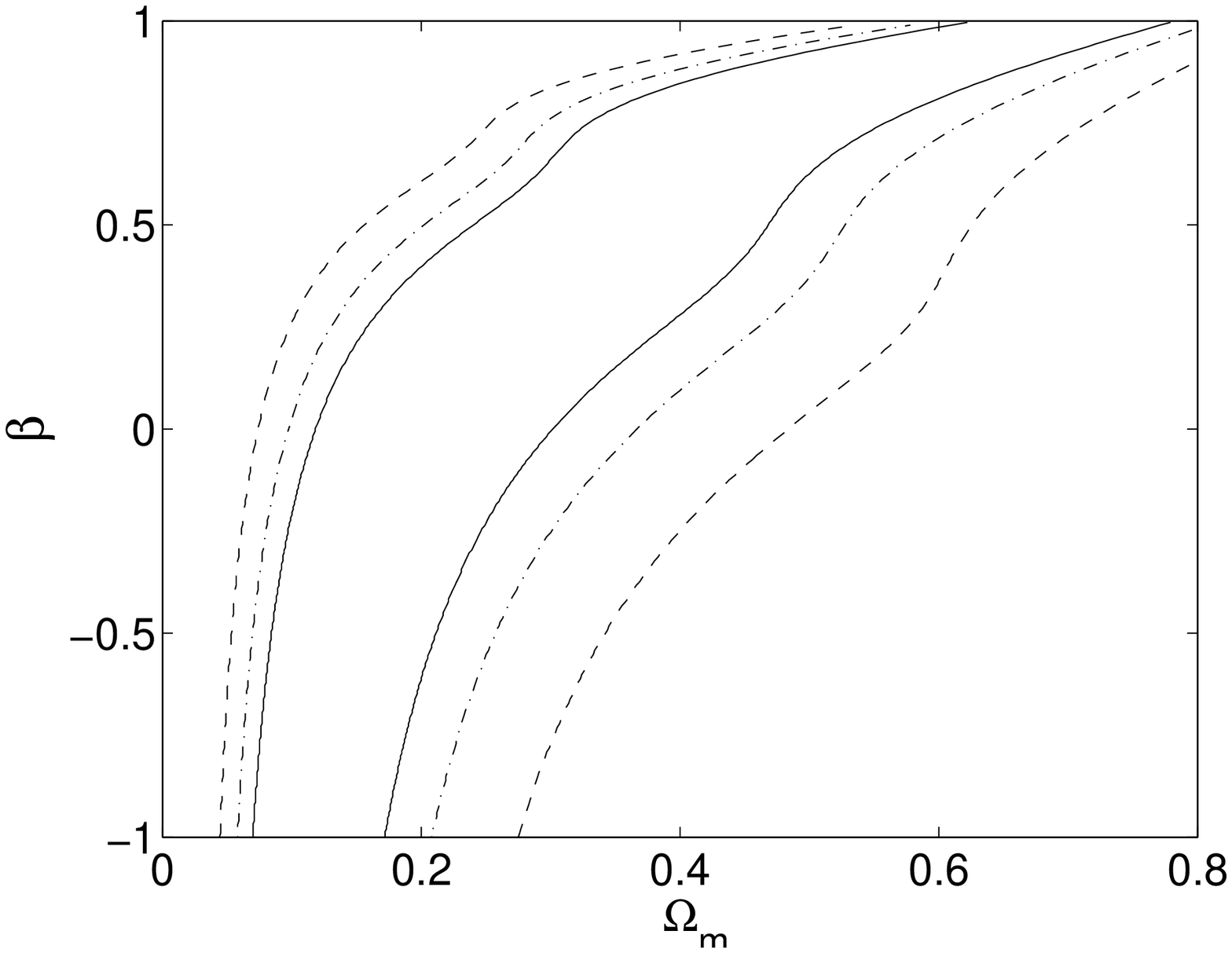}
  \includegraphics[height=.2725\textheight]{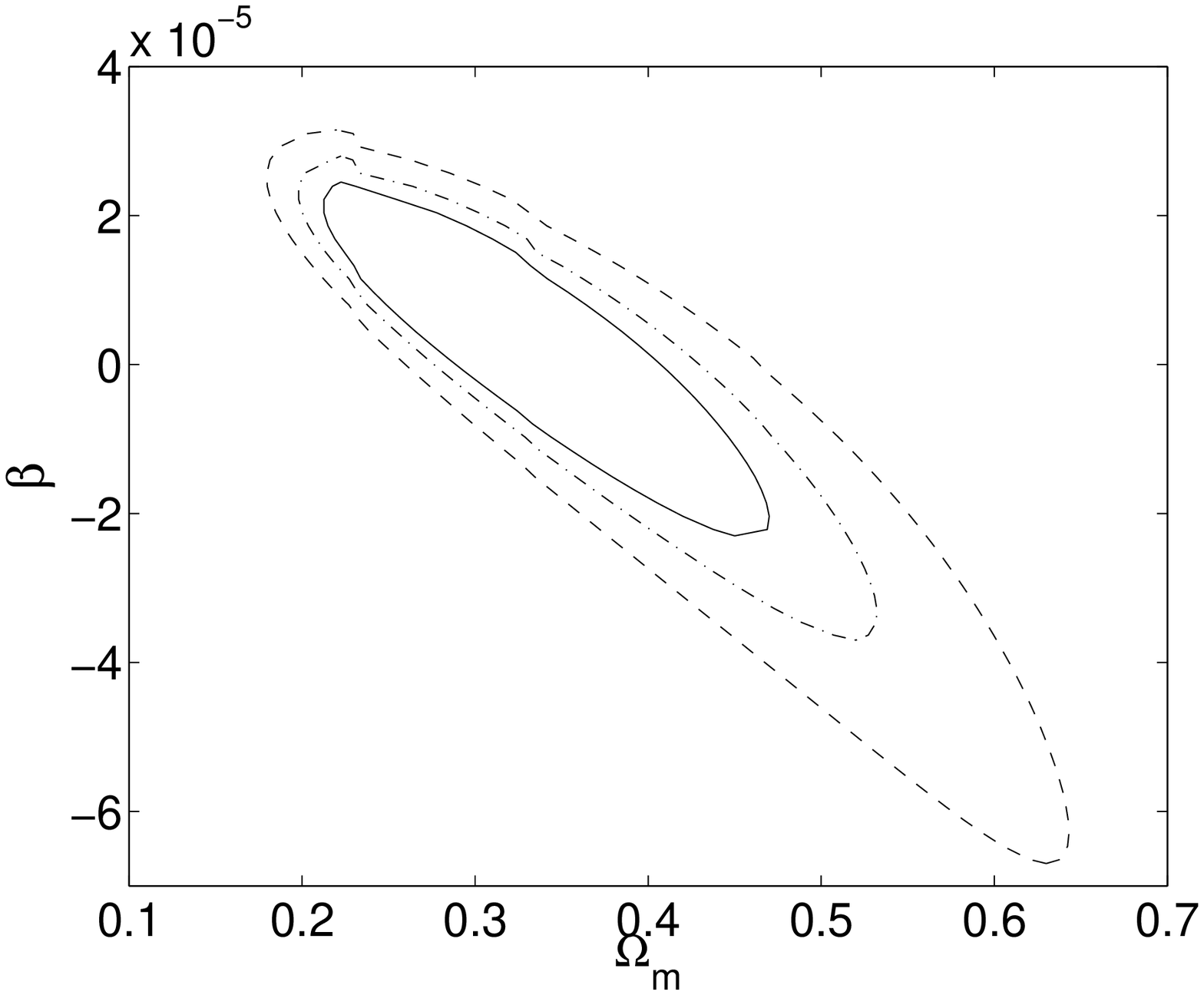}
  \caption{Constraints on the power-law Palatini model $f(R)=R+\alpha R^\beta$. Left panel shows the background constraints arising from from fitting the CMB shift parameter.
Right panel shows the perturbation constraints ensuing from fitting to the observed matter power spectrum. The contours correspond to 68, 90 and 99 \% confidence levels.
We allowed generous variation of the spectral index $n_S = 1 \pm 0.2$. Note that the $\beta$-axis of the right panel is scaled by $10^{-5}$. \label{palafig}}
\end{figure}

In Palatini-f(R) cosmology \cite{Vollick:2003aw,Barraco:2002xv}, the Friedmann equation obtained from (\ref{p_fe}) can be written as
\be
6F\left(H+\frac{1}{2}\frac{\dot{F}}{F}\right)^2 = \rho+3p - f\,.
\ee
If $w$ is constant this may be rewritten as \cite{Amarzguioui:2005zq}
\be
6H^2 = \frac{1}{(1-3w)F}\frac{(1+3w)RF-3(1+w)f}{\left(1-\frac{3}{2}(1+w)\frac{F(RF-2f)}{F(RF''-F)}\right)^2}\,.
\ee
We have expressed the Hubble rate as a function of $R$, which we in turn may solve from the trace
equation (\ref{trace}). Thus it is straighforward to find the scale factor evolution given the matter content.
It turns then out that these models can generate a viable sequence\footnote{A simple quadratic term can also generate a bouncing cosmology \cite{Barragan:2009sq}, showing these models might have relevance as an effective
description of loop quantum cosmology \cite{Olmo:2008nf}.} of radiation dominated, matter dominated and
accelerating era matching with the constraints. This has been shown for various parameterizations of the function $f(R)$, most often with some power-law forms
(with one, two or three powers of $R$), but also with square-root, logarithmic and exponential forms for the curvature correction
terms \cite{Capozziello:2004vh,Amarzguioui:2005zq, Borowiec:2006hk,Movahed:2007cs,Fay:2007gg}. As an example we show, in the left panel of Fig. \ref{palafig}, the constraints arising from fitting the 
the last scattering distance data to the prediction of the model defined by a monomial correction to the Einstein-Hilbert action.

At the Solar system scales, it seems that these models can correctly reproduce the GR phenomenology. Since the gravity correction in vacuum reduce to GR with a cosmological constant, and the
magnitude of the effective cosmological constant is given by the scale of acceleration today, the models can pass the local gravity tests. 

The problems occur at the level of inhomogeneous perturbations. For the background, the additional derivative terms in the field equations
due to nonlinearity in $f(R)$ can play the role of an effective smooth dark energy, but the spatial gradients of these extra terms cannot be neglected for
cosmological perturbations \cite{Koivisto:2005yc,Uddin:2007gj} which then assume unusual behaviour that is at odds with the observed distribution of galaxies \cite{Koivisto:2006ie} and with the
cosmic microwave background spectrum \cite{Li:2006vi}, even if the nonlinear part of the action is exponentially suppressed at late times \cite{Li:2006ag}. In quantitative terms,
parameterizing the $f(R)- R \sim R^\beta$, data analysis constrains  $|\beta| < 10^{-5}$ \cite{Koivisto:2005yc,Li:2006vi}. We show the confidence contours arising from
matching with SDSS data \cite{Tegmark:2003uf} in the right panel of Fig. \ref{palafig}.

However, there is way out of the problem. If we consider a dark matter component which is not entirely cold, but may have finite sonic, anisotropic or entropic stresses at large scales the effects of the
modified gravity on the matter power spectrum may be partially or completely eliminated. It is then possible to obtain exactly the same linear matter growth as in a universe with smooth dark energy producing the
same background expansion. Such Palatini models with generalized dark energy may be called $f(R)$GDM models instead of $f(R)$CDM models which do not are not viable alternatives to the $\Lambda$CDM models. The evolution of the dark matter overdensity is,
in general coupled to the evolution of the anisotropic stress $\Pi$ as 
\be
\ddot{\delta} =
D_1 H\dot{\delta} +
\left(D_2 H^2 + k^2 D_k\right) \delta +
P_1 H\dot{\Pi} +
P_2 H^2\Pi.
\ee
The exact form of the five dimensionless coefficients appearing in this equation can be read from the appendix of the Ref.\cite{Koivisto:2007sq}, where it was shown that any $f(R)$GDM satisfying the relation
(the sound speed $c^2$ can vanish as for CDM)
\be \label{shear}
\Pi = \frac{1}{1-3K/k^2}\left(\frac{\dot{F}}{4HF} + \frac{3}{2}c^2\right)\delta,
\ee
escapes the matter power spectrum constraints completely. However, due to presence of anisotropic stress now both in the matter and in the gravity sectors \cite{Koivisto:2005mm,Koivisto:2008ig}, the 
gravitational potentials will respond differently to the clumps of matter. These potentials affect the lensing experiments and also the CMB via the integrated Sachs-Wolfe effect \cite{Zhang:2007nk}. Thus one may distisguish
$f(R)$GDM from $\Lambda$CDM by exploiting these data. The former models have not yet been confronted with the data. A recent data analysis using a general parameterization \cite{Bean:2009wj} favors 
standard growth rate but nonzero effective anisotropic stress, which is the generic $f(R)$GDM prediction.     


\section{Nonlocal models}
\label{nl-sec}

A class of nonlocal gravity models can be defined by the action 
\be 
S  = \int d^4x\ \sqrt{-g}\left( R + 
\sum_{n=0}^{\infty}{\hat{c}_{n}}R\Box^{n}R \right) \, .
\label{non-pert}
\ee
Thus the model is specified by the infinite number of coefficients $\hat{c}_n$. One could also write them as
$\hat{c}_n=c_n/M_\ast^{2(n+1)}$, where $c_n$ are dimensionless and if roughly of order one,
$M_{\ast}$ is the mass scale at which the higher derivative terms in the action become important. 
To avoid the generic higher-derivative problems, the coefficients $c_n$ must remain non-vanishing to arbitrarily high orders. 
This string-inspired form of nonperturbative gravity has been also shown to be asymptotically free and thereby 
potentially describe bouncing cosmologies \cite{Biswas:2005qr,Biswas:2006bs}. 
Similar models may be also motivated by quantum loop corrections to gravity.
We consider then the following simple example\footnote{For cosmology of nonlocal scalar field models motivated by string field theory see 
for example \cite{Aref'eva:2004qr,Aref'eva:2007uk,Joukovskaya:2007nq,Koshelev:2009ty}.}
of a nonlocal action \cite{Deser:2007jk, Nojiri:2007uq}
\be 
S  =  \int d^4 x \sqrt{-g}\left[ R\left(1+f(\Box^{-1}R)\right) + 2\kappa^2\mathcal{L}_m \right], \label{eka}
\ee
where $f$ can be called the nonlocal distortion function.
If we consider analytic functions $f(x)$, 
\be \label{f_ans}
f(x)=\sum_{i=0}^\infty f_i x^i\,,
\ee
then the actions can be shown to reduce, at least patchwise, within the class (\ref{non-pert}). 
Consider the simple case $f(x)=f_1 x$. Such correction was proposed to bound the Euclidean action and its cosmological effects were considered already in Ref.\cite{Wetterich:1997bz}.
This model is equivalent to the series specified by
\be
c_n = \frac{(-1)^n}{n!}f_1 \sum_{k=0}^n\frac{k!}{(n-k)!} \,.
\ee
One should note here that the scale $\Ma$ must be such that $-\Box R \ll \Ma^2 R$. In a similar way it is straightforward to see
that with the same caveat, any analytic $f$ can be presented via such a power series (the coefficients $c_i$ are then given by a double sum whose explicit form is not needed here).

Introducing the Lagrange multiplier $\xi$ to rename the inverse d'Alembertian as $\phi$, we may rewrite the action in the form
\be \label{act0}
S  =  \int d^4 x \sqrt{-g}\left[R\left(1+f(\phi)\right) + \xi(\Box\phi -R) + 2\kappa^2\mathcal{L}_m\right].
\ee
This is a peculiar kind of double-scalar-tensor theory where massless, nonminimally coupled scalars have a kinetic-type interaction resembling
of the nonlinear sigma model. 
If we then define 
\be
\Psi(\phi,\xi) \equiv f(\phi) - \xi,
\ee
integrate partially and neglect the boundary terms as usual, it 
it becomes transparent that we may consider a biscalar-tensor theory, with only one of the scalar degrees of freedom is nonminimally coupled to gravity as
\be \label{act2}
S =  \int d^4 x \sqrt{-g}\left[(1+\Psi)R -
f'(\phi)(\partial\phi)^2 + \nabla_\alpha\Psi\nabla^\alpha\phi + 2\kappa^2\mathcal{L}_m\right] \,.
\ee
However, like in reducing the action into the form (\ref{non-pert}), we should note some caveats in reducing it into the form (\ref{act2}). Namely,
when getting rid of the inverse d'Alembertian in step (\ref{act0}), we are introducing a second order constraint into the Lagrangian. In general,
this is not legitimate, as pointed out by Koshelev \cite{Koshelev:2008ie}. The solution to the constraint equation $\Box \phi = R$ is unique only up to 
the most general harmonic function $\varphi$ for which $\Box \varphi =0$ in a given background. Thus the local form of the model seems to introduce a new degree of freedom which the original theory lacks:
the Lagrangian multiplier becomes a dynamical variable. From the practical point of view of finding classical solutions the relevant question is whether one can recover the solutions of 
the original theory by just choosing the initial conditions suitably, i.e. choosing the correct function $\varphi = 0$. As a particular case, consider the limit $f \rightarrow 0$. The original
theory obviously reduces to GR, while the spurious degrees of freedom remain in the action (\ref{act2}): however, it is clear that the latter are not actually there if we impose the correct boundary condition 
for the equation of motion\footnote{In this simplest (and trivial) case, the equation of motion for $\Psi$ is $\Box\Psi=0$ and the correct boundary condition reproducing the original model is $\Psi=0$.}. 
This is the only case where the non-equivalence has been explicitly shown \cite{Koshelev:2008ie}, but a more general treatment of the issue would be worthwhile, especially since the local form
(\ref{act2}) (with the prescription for initial conditions understood) is considerably more tractable than the manifestly nonlocal starting point (\ref{eka}). A rigorous mathematical analysis of the 
possible equivalence\footnote{Recent localization considerations \cite{Barnaby:2008tc,Calcagni:2007ru,Vernov:2009vf} have for example employed reformulation of the nonlocal system as diffusion 
problem \cite{Mulryne:2008iq,Calcagni:2009jb}.} is
outside the scope of present discussion, and thus one should regard some the following results strictly to apply only to the {\it local variant} (\ref{act2}) of the model (\ref{eka}).    
  
\subsection{Nonlocal models: Cosmology}

The reasons to suspect a correction involving $R/\Box$ as a culprit for dark energy can be easily seen. This term is dimensionless. Thus, one may construct the distortion
function without introducing any dimensional functions. It appears as a prefactor for the Newton's constant, and explain weakening of gravity at cosmological scales, the 
distortion function should be about minus few dozens of percents today. Thus no new scale needs to be introduced, in particular the hierarchy between the observed magnitude 
of dark energy density and the Planck scale does not to have to prescribed into the action. Another side of the problem is then why the correction is important at the present 
stage, and in this model the explanation is that during the radiation dominated era the Ricci scalar is vanishing. So the evolution of the distortion function and the ensuing 
acceleration is a consequence of the universe entering into the matter dominated stage. -The possible problems are also easily seen. Namely, it is not clear that the distortion effects would be confined 
solely to the cosmological scales and the question arises why there aren't observed elsewhere. 

Analytic and numerical solutions for the background expansion in this model in a realistic universe were found in \cite{Koivisto:2008xfa}, see also \cite{Jhingan:2008ym,Capozziello:2008gu,Cognola:2009jx}.
A method to reconstruct the model generating an arbitrary background history was also constructed for both the scalar-tensor form and the original model \cite{Deffayet:2009ca}. 
Another way to proceed is to specify a function $f$ and compute its predictions for the FRW metric. For simplicity, consider the monomial $f(x)=f_n x^n$ (no summation). We may then schematically summarize the 
possible evolution histories as follows:
\begin{displaymath}
\xymatrix{
f=f_n\phi^n \ar[r]^{n>0} \ar[d]_{n<0} & \text{Nonlocal effect} \ar[d]_{(-1)^{n}f_n>0} \ar[dr]^{(-1)^{n}f_n<0} &     & \\
\text{Matter domination}                                & \text{Slows down expansion} \ar[l] & \text{Acceleration} \ar[r] & \text{Singularity}
}
\end{displaymath}
Thus, with positive the power $n$ and the sign $f_n$ equal to\footnote{In fact this sign seems necessary to avoid a ghost.} $-(-1)^n$, one finds dark energy solutions. This scenario leads to a sudden future singularity, which might be avoided be regularization of the
operator or reshaping the function $f$. Though this was not in the focus of the present study, $n<0$ may lead to a decaying effect which, with large enough scale $f_n$, can influence
the evolution of the early universe. 

The linear inhomogeneity field of matter perturbed around the metric (\ref{metric}) evolves as \cite{Koivisto:2008dh} 
\be \label{d_evol3}
\ddot{\delta} + 2H\dot{\delta} = 4\pi G_* \rho_m\delta =
\frac{1}{2}\left(\frac{(1+\psi-8f')\kappa^2}{(1+\psi-6f')(1+\psi)}\right)\rho_m\delta_m\,.
\ee
The brackets in the last expression enclose the effective gravity coupling. We see it is time-dependent but scale-invariant. We then do get nontrivial modifications of the growth rate, but
don't find the problematic effects of a nonzero sound speed like in the first order gravity models of section \ref{pala_struc}. The time-dependence of the gravitational constant can in principle be used 
to distinguish a nonlocal origin of acceleration, when the background expansion is identical to some local model.  

The perturbations about the Schwarzchild geometry for the model (\ref{f_ans}) can be shown to imply the post-Newtonian parameters
\be
G_* = \left(\frac{1+f_0-8f_1}{1+f_0-6f_1}\right)\frac{G_N}{1+f_0}, \quad \gamma = \left(\frac{1+f_0-4f_1}{1+f_0-8f_1}\right).
\ee
Therefore it is possible to constrain the constant and the linear parts of the coupling from Solar System experiments. However, the constant $f_0$ may always
be absorbed into a redefinition of the $\kappa \rightarrow \kappa/(1+f_0)$, and so one may set $f_0=0$. We then get the strict bound
$-5.8 \cdot 10^{-6} < f_1 < 5.7 \cdot 10^{-6}$. Deviations from GR ensuing from the higher order terms in the expansion for $f(\phi)$ are of the second order in perturbations, 
$\sim (G_NM/r)^2$, and thus seem undetectable within present experimental accuracy. However, we have only shown that such solutions exist. Whether the assumption that we may expand 
about the Schwarzchild geometry is realistic should be assessed with more care. Properties of exact spherically symmetric solutions were studied in \cite{Bronnikov:2009az}.

\subsubsection{Nonperturbative gravity and holographic cosmology} 

It is interesting to remark that nonperturbative string-inspired (\ref{non-pert}) or quantum loop corrected (\ref{eka}) gravity models
could provide a covariant realization of the idea of holographic cosmology. The latter is indeed some manifestation of nonlocal gravity. As we explained above, the nonlocal distortion function can
automatically feature effects at the right energy scale, thus alleviating the cosmological fine-tuning problems. A similar promising feature can be found in the holographic dark energy models. The holographic 
principle, familiar from black holes \cite{Bekenstein:1973ur} and quantum gravity \cite{'tHooft:1993gx}, is postulated to set an upper bound on the 
entropy of the universe \cite{Fischler:1998st}. The quantum zero point energy of the region of size $L$ should not then exceed the mass of a black hole of the same size, $L^3\rho_\Lambda \le L/\kappa^4$. This then implies 
dark energy density scaling as $L^{-2}$, and if $L$ is proportional to the Hubble horizon $L \sim 1/H$, the fact that dark energy density today happens to be of the observed magnitude becomes almost tautological. 
However, the details do not work correctly unless the horizon is given by the future event horizon \cite{Li:2004rb},
\be \label{event}
L(t) = a(t)\int_t^{\infty}\frac{dt'}{a(t')} = a(t)\int_{a(t)}^{\infty}\frac{da}{H(a)a^2}\,.
\ee    
This is the boundary of the volume an observer at time $t$ may eventually see. At this level, we only have an arguably strange modification of the Friedmann equation which lacks a covariant justification. However,
it is not unconceivable that similar cosmologies may be derivable from a nonlocal theory of the form (\ref{non-pert}) or its variants. In particular, if one takes the action (\ref{eka}) and considers the inverse 
d'Alembertian to correspond to, instead of the retarded Green function, the advanced one, then future integrals like those in the expression for the event horizon (\ref{event}), appear into the effective field 
equations. For example, in FLRW background we have
\be
\frac{1}{\Box}\Phi(t) = - \int_{t}^{\infty}\frac{dt'}{a^3(t')}\int_{t'}^\infty dt''a^3(t'')\Phi(t'')\,.
\ee
Producing exactly the modification (\ref{event}) seems difficult\footnote{By this we mean the manifest form of the field equations. Of course we can reconstruct models with identical background expansion to the
one implied by the phenomenological Friedmann equation involving the integrals (\ref{event}).}
but is probably unnecessary, as the main aim would be to produce dynamical dark energy whose magnitude would depend upon some suitably defined
horizon size.

\section{Conclusions}
\label{co-sec}

Modified gravity often contains pathological degrees of freedom originating from higher derivatives or extra fields. Avoiding introduction of new propagating modes is possible in some special cases, whose cosmological
implications were considered here. In the framework of scalar-tensor theories, a special one-parameter family of theories is formed by those in which the scalar field is an algebraic function of the Ricci scalar 
and the trace of the matter tensor. In two limits these give these models reduce to the $f(R)$ models, in the metric and in the Palatini formulation. In addition, one may consider coupling the scalar to the Gauss-Bonnet
invariant. Finally, by considering nonlocal modifications one may avoid the appearance of a finite number of poles into the propagator, whilst (multi)scalar-tensor modeles may 
reflect some properties of these theories. 

The Gauss-Bonnet model can feature a transition from scaling era to dark energy domination, but there are problems with consistent solar system tests and new instabilities at cosmological scales.
The Palatini models may viably reproduce the observations at both the cosmological and smaller scales, if one assumes some non-cold properties for the dark matter component.
The definite predictions of the nonlocal models are presently less robustly understood, but further study of such models is worthwhile as they seem promising in alleviating the cosmological fine-tuning problems.    


\begin{theacknowledgments}

I thank David F. Mota for discussions and collaborations. 

\end{theacknowledgments}



\bibliographystyle{aipproc}   


\IfFileExists{\jobname.bbl}{}
 {\typeout{}
  \typeout{******************************************}
  \typeout{** Please run "bibtex \jobname" to optain}
  \typeout{** the bibliography and then re-run LaTeX}
  \typeout{** twice to fix the references!}
  \typeout{******************************************}
  \typeout{}
 }

\bibliography{etgrefs}

\end{document}